\documentclass{article}
\RequirePackage{filecontents}
\PassOptionsToPackage{dvipsnames}{xcolor}
\RequirePackage{comgipp}
\RequirePackage{amsfonts}
\RequirePackage{amsmath}
\RequirePackage{authblk}

\newcommand{\theReferenceText}{\fullcite{GreinerPetter2018b}}
\usepackage[
backend=biber,
style=numeric,
firstinits=true,
url=false,
isbn=false,
safeinputenc,%
]{biblatex}
\title{Automatic Mathematical Information Retrieval to Perform Translations up to Computer Algebra Systems}

\author[1]{Andr\'{e} Greiner-Petter}
\affil[1]{University of Konstanz, Konstanz, Germany\\ \texttt{andre.greiner-petter@t-online.de}}

\date{January 2019}
\begin{document}
  \maketitle
  \thispagestyle{firststyle}
  \begin{abstract}
    In mathematics, \LaTeX{} is the de facto standard to prepare documents, e.g., scientific publications. While some formulae are still developed using pen and paper, more complicated mathematical expressions used more and more often with computer algebra systems. Mathematical expressions are often manually transcribed to computer algebra systems. The goal of my doctoral thesis is to improve the efficiency of this workflow. My envisioned method will automatically semantically enrich mathematical expressions so that they can be imported to computer algebra systems and other systems which can take advantage of the semantics, such as search engines or automatic plagiarism detection systems. These imports should preserve essential semantic features of the expression.

  \end{abstract}
\section{Problem \& Motivation}
The general problem of enriching mathematical expressions with semantic information and providing lossless translations to computer algebra systems can be divided into two parts: the translation process and the semantic enrichment process, hereafter called semantification. The first part was the focus of my Master\'{}s thesis (see section~\ref{sec:MA}). The focus of my doctoral research will lie on the second part of the problem. Providing extractions automatically for semantic information of mathematical expressions is worthwhile for several different tasks and can be examined from different areas of interests. For example, semantic knowledge strongly improves mathematical search engines~\cite{SemanticSearch,PHD:Moritz}. Imagine someone has the famous Pythagorean theorem with different variable names, e.g. $q^2+p^2=r^2$. Finding a connection between this formula and the Pythagorean theorem is still a difficult task for nowadays search engines. Another, rather new field of interest is automatic plagiarism detection in mathematical expressions~\cite{Plagiarism}. Detecting plagiarism in a mathematical formula is highly imprecise, if the semantics of the formula is unknown. As seen in the example above, finding the connection between $q^2+p^2=r^2$ and the Pythagorean theorem probably remains unrecognized, which makes it difficult to see a potential plagiarism in this formula.

During this extended abstract, I will explain the difficulties of an automatic semantification process on examples related to problems we have for converting mathematical expressions to computer algebra systems. In the following, I will give a brief introduction to this workflow and explain its issues in a more detailed way.

Scientists usually work with word processors to write scientific papers. The very well-known word processor \LaTeX\footnote{Since \LaTeX{} is more than just a word processor, calling \LaTeX{} that way might not be accurate enough. But during our research, we compared the theoretical concepts behind \LaTeX{} and word processors such as Microsoft Word and we decided to keep this description.} has become a de facto standard\footnote{\url{https://www.latex-project.org/}} for this purpose over the last 30 years~\cite{Latex:Standard}. Also, numerous other editors, such as the editor for Wikipedia articles or Microsoft Word, entirely or partially support \LaTeX{} expressions. \LaTeX, developed by Leslie Lamport, extends the typesetting system \TeX, developed by Donald E. Knuth 1977~\cite[p. 559]{Knuth}, by a set of standard macros. Knuth created the \TeX{} system because he was unsatisfied with the typography of his book, The Art of Computer Programming~\cite[pp. 5-6, 24]{Knuth}. \LaTeX{} and \TeX{} provide a syntax for printing mathematical formulae in a way a person would write it by hand. During my research, I will primarily focus on \LaTeX{} formats.

\begin{table}[t]
\centering
\small
\begin{tabular}{cl}
	\hline
	Systems & \hspace*{2cm}Representations\\
	\hline
	Rendered Version & $P_n^{(\alpha, \beta)}(\cos(a\Theta))$\\
	Generic \LaTeX & \verb|P_n^{(\alpha, \beta)}(\cos(a\Theta))|\\
	Semantic \LaTeX & \verb|\JacobiP{\alpha}{\beta}{n}@{\cos@{a\Theta}}|\\
	CAS Maple & \verb|JacobiP(n, alpha, beta, cos(a Theta))|\\
	CAS Mathematica & \verb|JacobiP[n,\[Alpha],\[Beta],Cos[a\[CapitalTheta]]]|\\
	\hline
\end{tabular}
\caption{Example of different representations for a Jacobi polynomial}
\label{tab:useCase}
\end{table}

Besides that, scientists work with formulae in their papers while they evaluate special values, create diagrams and find or calculate practical solutions. Computer algebra systems (CAS) are software tools that allow such computations on mathematical expressions. Since CAS should be easy to use, they created their representations with the intent of displaying the results as intuitively as possible. That approach for representing formulae is not standardized. That is why most CAS have created their representation, all of which are different.

Mathematical expressions written in \LaTeX, which use only standard libraries of macros (hereafter called generic \LaTeX{}), do not provide sufficient semantic information about the used symbols. In contrast to that, a certain level of semantic information is necessary for CAS to offer correct computations on input expressions. The exact semantics concluded by readers is based on their knowledge and in which context the expression is used. Without a translation tool, a typical workflow for scientists contains different representations for the same mathematical expression: a representation in a word processor such as \LaTeX, and another representation in a CAS such as Maple or Mathematica. Table~\ref{tab:useCase} in section~\ref{sec:relatedWork} describes different representations for the same mathematical expression.

The semantic enrichment problem can be made clear with the description of a single symbol expression. Consider the Euler-Mascheroni constant represented by the Greek letter $\gamma$. Without any further information, $\gamma$ is just a Greek letter, often used to represent this mathematical constant, but can also be used to represent curve parametrization, and many other things, such as a variable. In \LaTeX, $\gamma$ is represented as \verb|\gamma|. The equivalent representation in a CAS, such as Maple or Mathematica, depends on the meaning of $\gamma$. For example, in Maple and Mathematica, the Euler-Mascheroni constant is represented as \verb|gamma| and \verb|EulerGamma| respectively. Another more complicated example might be Euler\'{}s number e, which is well-known to be the constant used for the basis of the natural logarithm. Translated to Mathematica, it is just a capital \verb|E|, while Maple has no specific symbol set aside to represent the constant, and one has to evaluate the exponential function using \verb|exp(1)|. 

The above examples indicate two potential problems. A translation with semantic information might be difficult because one needs to be aware of the details of the CAS. For instance, a scientist who usually works with Mathematica might not know that \verb|gamma| in Maple is not only a Greek letter but also the constant. Therefore, an automated equivalent translation between representations is desirable. The other problem occurs when the semantic information is not sufficient or is essentially absent in the expression. For example, consider the mass-energy equivalence formula $E=mc^2$. In plain \LaTeX{} \verb|E|, \verb|m| and \verb|c| are Latin letters and one has to analyze the context of each symbol to pick the correct semantic information. It is essential to clarify the correct semantics before translating an expression to another representation. Otherwise, \verb|E| might be mistakenly translated as a Latin letter to "\verb|E|" in Mathematica instead of \verb|EE| to represent the energy. 

\section{State of the Art}\label{sec:relatedWork}
Unfortunately, there is no agreement on how much semantic information is necessary for an expression to be sufficiently semantic. It always depends on the task. For example, in most cases, no semantic information is necessary to render a mathematical expression. For a search engine, providing the name of the formula might be adequate. But for computations on mathematical expressions, detailed knowledge about the definitions becomes necessary. In this context, we suggest that a mathematical expression is sufficiently semantic when a translation to a CAS becomes feasible.

Before we can start a semantification process, we need a system that is capable of carrying such information. There are several approaches for attaching semantic information to symbols or entire expressions. One approach is the content Mathematical Markup Language (content MathML)~\cite{MML}, which tries to organize semantic information in an XML document. MathML and content MathML are widely used for web services because of their simple and easy to parse XML structure. The JOBAD architecture~\cite{JOBAD,CICM10:Jobad} uses content MathML to create web documents with access to semantic information. While XML is an appropriate format for algorithms, it is not convenient for humans. 

The National Institute of Standards and Technology (NIST) in Maryland, USA, has developed a set of \LaTeX{} macros to tie specific character sequences to well-defined mathematical objects and thereby providing semantic information within \LaTeX{} expressions. NIST uses these macros for the Digital Library of Mathematical Functions (DLMF)~\cite{DLMF} and the Digital Repository of Mathematical Formulae (DRMF)~\cite{DRMF,DRMF2}, an outgrowth of the DLMF project. Thereby we call these set of macros \textit{DLMF/DRMF macros}.

\begin{figure*}[t]
	\includegraphics[width=\textwidth]{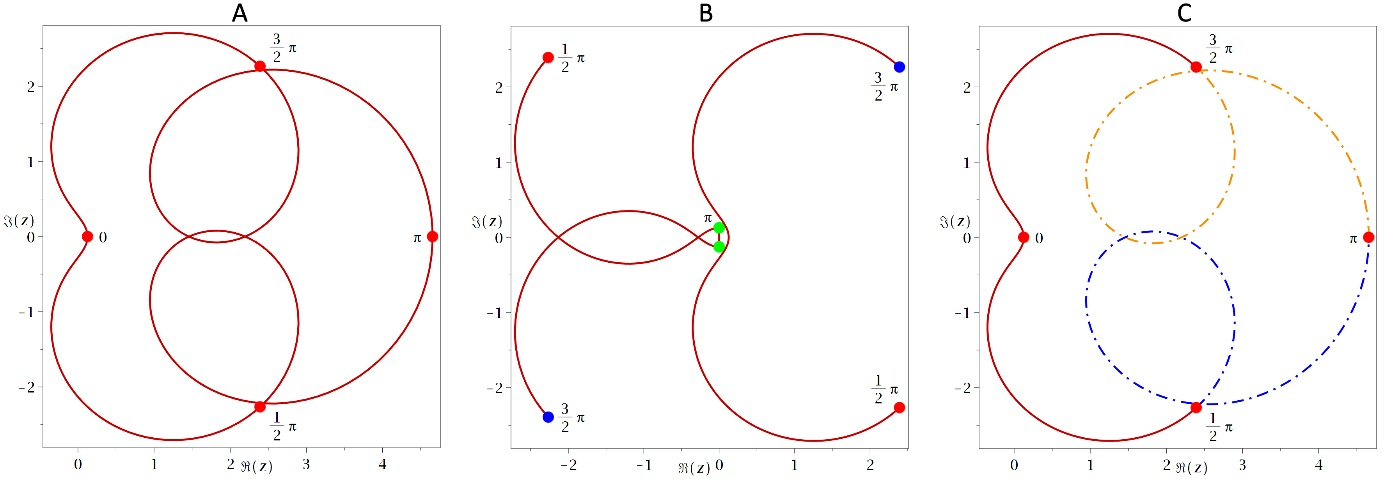}
	\caption{Polar plot for the parabolic cylinder function with $z\left( \phi \right) = 2.5e^{i\phi}$ for $\phi \in \left[0,2\pi\right]$. Subfigure A shows $U\left(0,z\left(\phi\right)\right)$. Subfigure B shows the right-hand side of DLMF 12.7.10. Subfigure C uses analytic continuation to allow computations on other branches.}
	\label{fig:branchCutProblems}
\end{figure*}

Table~\ref{tab:useCase} gives an overview of several different representations with and without semantic information for a Jacobi polynomial. The rendered version illustrates how a scientist would write the expression by hand. As previously explained, the generic \LaTeX{} expression and the rendered version does not provide sufficient semantic information in the source, while the DLMF/DRMF macro, Maple and Mathematica representations are tied to specific definitions.

Besides the noticeable visible differences in used parentheses or the ordering of the arguments, more complicated and hidden differences may appear for multi-valued functions. CAS usually define branch cuts to compute principle values for multi-valued functions. Thereby, CAS implementations of multi-valued functions are discontinuous. Moreover, the position of these branch cuts varies from CAS to CAS~\cite{BranchCut}. It is important that CAS users understand the position of such cuts~\cite{MAPLE:BranchCut}. For example, the parabolic cylinder function $U(a,z)$ is defined as a solution of a second order differential equation\footnote{\url{http://dlmf.nist.gov/12.2\#SS1.p1}}. $U(a,z)$ can also be represented by other functions, such as the modified Bessel function of the second kind $K_\nu(z)$. One relation between $U$ and $K_\nu$ is:

\begin{equation}
U(0,z) = \sqrt{\frac{z}{2\pi}} K_{\frac{1}{4}}\left(\frac{1}{4}z^2\right), \quad \text{for } z \in \mathbb{C}.
\end{equation}

Figure~\ref{fig:branchCutProblems} visualizes the problem of this relation using polar plots, for fixed $r=2.5$ in the polar representation of complex numbers $z=re^{i\phi}$. Figure~\ref{fig:branchCutProblems}.A draws a curve for $0\leq\phi\leq2\pi$ of $U\left(0,z\left(\phi\right)\right)$. The curve is continuous and closed. Figure~\ref{fig:branchCutProblems}.B draws the same curve for the principal values of the right-hand side of the equation as computed by a CAS. The function jumps over branch cuts at the angles $\phi\in \lbrace \frac{\pi}{2}, \pi, \frac{3\pi}{2} \rbrace$. Based on the relation above, Figure~\ref{fig:branchCutProblems}.B should yield to the same curve as Figure~\ref{fig:branchCutProblems}.A. The solution for this problem is analytic continuation  which allows computations on other branches. Figure~\ref{fig:branchCutProblems}.C shows the right-hand side of the equation with analytic continuation for $K_\nu(z)$. The orange $\pi\leq\phi \leq\frac{3}{2}\pi$ and blue $\frac{1}{2}\pi \leq \phi \leq \pi$ curves visualize the respective branches, computed with analytic continuation, while the red $0 \leq \phi \leq \frac{1}{2}\pi$, $\frac{3}{2}\pi \leq\phi \leq 2\pi$ curve is the principal value. If CAS users use B rather than C for their computations, this leads to incorrect conclusions.

Most CAS provide import and export functions to other representations (see~\cite{MAPLE:Import,MATHEMATICA:Import,MA}). However, most import functions are only provided for presentations that already carry semantic information, foremost content MathML. While import and export functions rarely take care of system-specific differences~\cite{MA}, such as in case of different branch cut definitions from the example above, the problems become more serious when the semantic in the representation is unclear. In this case, the tool has to be able to extract the semantic information first to allow a translation. To my knowledge, there is no translation tool available that solves this problem. A typical workaround is to use translation tools, that can transform mathematical \LaTeX{} to MathML formats. These tools also contain a lot of problems and cannot handle the mentioned hidden differences. We already studied those tools and their accuracy in a paper~\cite{JCDL:Paper} (see section~\ref{sec:extractSemantic}).

However, finding ways of an automatic semantification process is an active topic of research. Nowadays, most semantification algorithms focusing on natural languages, which leads to the well-studied and broad field of natural language processing (NLP). In this respect, a relatively new field of interest is to extend this research for mathematical expressions. Existing approaches (and work in progress projects) try to adapt the approaches from NLP~\cite{Moritz:Semant,MathWriting,MCAT,SemanticSearch,POM,DRMF2,MLP,MLP:Improvement,PHD:Moritz,JCDL:Paper}. Why this adaption process can be error-prone will be shown by an example I discovered during my research. Our test formula\footnote{\url{https://dlmf.nist.gov/1.5\#E2}} explains the continuity of a two-valued function $f$:

\begin{equation}
|f(a+\alpha,b+\beta) - f(a,b)| < \epsilon.
\end{equation}

The surrounding text contains the following part: "\textit{[$\ldots$] for every arbitrarily small positive constant $\epsilon$ [$\ldots$]}". Basically, every preceding and following noun of an identifier (in this case $\epsilon$) is a candidate for a definiens\footnote{Latin for a term that describes another term (in our case symbols or identifiers).}. We were using the Stanford CoreNLP\footnote{\url{https://stanfordnlp.github.io/CoreNLP/}} to tag words. Unfortunately, the tagger tagged the word "\textit{constant}" as an adjective rather than a noun, which leads to the wrong assumption that $\epsilon$ has no definiens in the surrounding text. While this example was an extreme case, the context analyzation process is usually time-consuming and mostly finds a lot of possible but not correct pairs of identifiers and definiens, i.e., a high recall\footnote{Fraction of relevant instances that have been retrieved from the sources over the total amount of relevant instances in the sources.} and a low precision\footnote{Fraction of relevant instances among the retrieved instances.} value. Note that a high recall value does not mean there would be no space for improvements, seen in the $\epsilon$-example above.  

\section{Preliminary Research}
My preliminary research is structured in two parts. The research starts with my Master\'{}s thesis, which I describe in the following subsection~\ref{sec:MA}. The second part focused on analyzing existing tools, creating a comprehensive gold standard and develope improvement techniques, explained in subsection~\ref{sec:extractSemantic}.

\subsection{Translation from semantic \LaTeX{} to CAS}\label{sec:MA}
\begin{figure}[t]
	\centering
	\includegraphics[trim=3 4 3 3,clip,width=\textwidth]{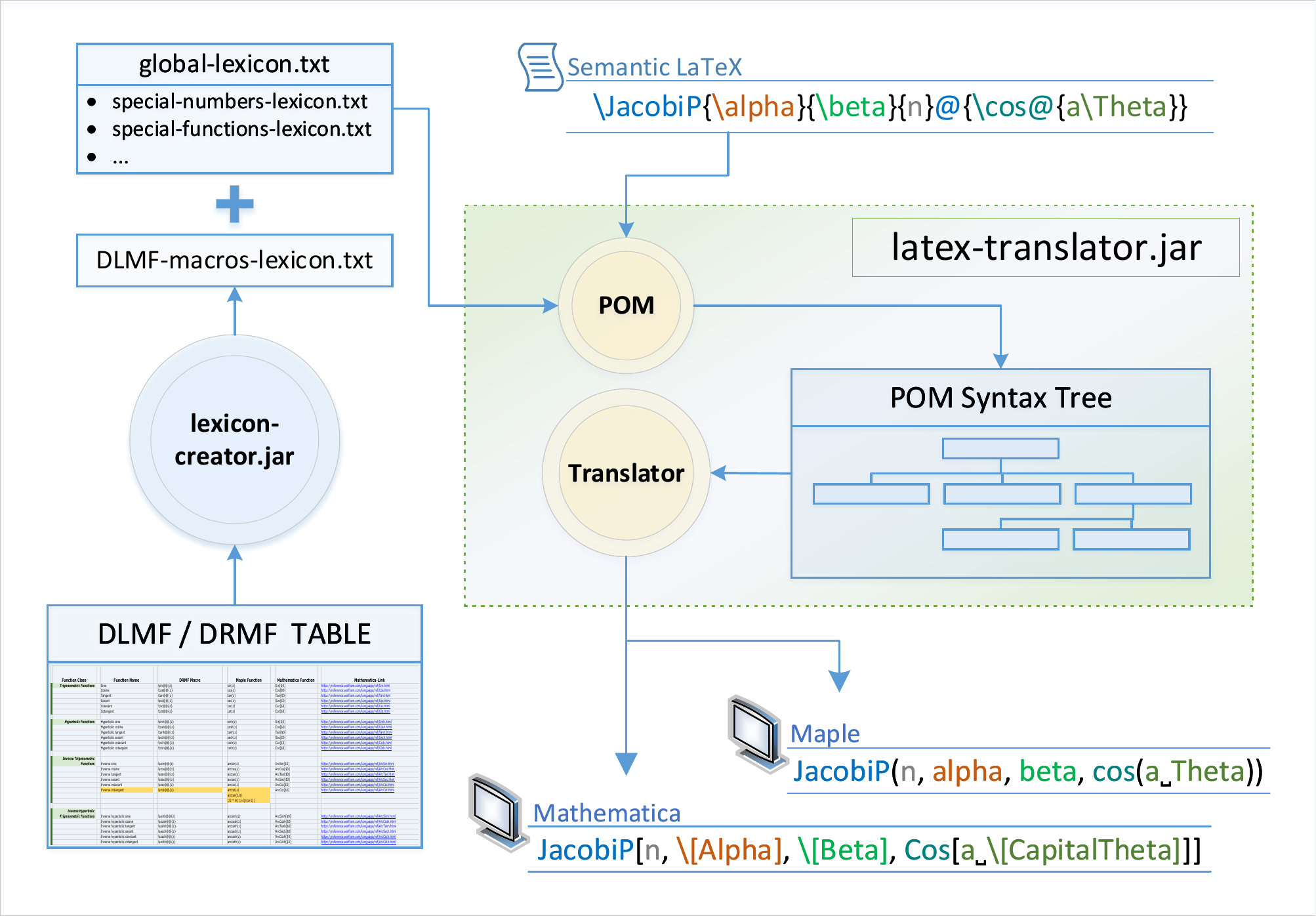}
	\caption{Flow diagram, which explains the translation process between semantic \LaTeX{} and a CAS. The POM-Tagger parses a \LaTeX{} expression based on the lexicon files and creates a syntax tree representation. That syntax tree can be translated node by node to a CAS representation.}
	\label{fig:transWorkFlow}
\end{figure}
I completed my Master\'{}s thesis at the TU Berlin in Germany, NIST, USA, and Maplesoft, Canada. I focused my research on the translation between special function and orthogonal polynomial \LaTeX{} expressions with DLMF/DRMF macros and representations in the Maple and Mathematica CAS. The thesis was also published in a conference paper~\cite{MA}.

From a scientific point of view, mathematics in \LaTeX{} is primarily a formal language with an alphabet, words, and rules. \LaTeX{} mathematical expressions can be described by a context-free grammar, to build a syntactical structure. We used the Part-of-Math tagger (POM-Tagger)~\cite{POM}, to build a syntax tree of a mathematical expression written in \LaTeX. With a syntax tree, we can translate an expression, node by node without changing the meaning of the expression.

Figure~\ref{fig:transWorkFlow} describes the translation process for the Jacobi polynomial example in table~\ref{tab:useCase}. The POM-Tagger uses the lexicon files to tag each token in the input expression with additional information. These lexicon files store that information, that describes the meaning of each symbol. I extended the lexicon files to provide translation patterns for the semantic DLMF/DRMF macros. Thereby, the POM-Tagger provides translation patterns for known semantic macros and also semantic but ambiguous information about other symbols. For instance, in the lexicon file, the symbol \verb|\beta| indicates that it is a Greek letter and it is often used to describe the second of three angles in a triangle, or in case of special functions, it could be the Dirichlet beta function, and so on and so forth. In our example from table~\ref{tab:useCase}, it is just a variable. Picking the right meaning is essential and one of the goals of my doctoral thesis.

For the Master\'{}s thesis, we only allowed disambiguated expressions with DLMF/DRMF macros, to avoid all of the problems previously described. We also followed the intuitive approach and presumed that the semantic information is already provided. Consequently, an author has to provide this information during the writing process. In my Master\'{}s thesis, I attempted to provide mathematical expression representations with sufficient semantic information to provide a lossless translation to a CAS. Therefore, we refer to a \LaTeX{} expression which uses DLMF/DRMF macros as semantic \LaTeX. 

\subsection{Extract semantics from generic \LaTeX}\label{sec:extractSemantic}
While the POM-Tagger follows a lexicon-based approach to tag tokens with semantic information, other approaches already try to analyze the context, i.e., the surrounding text, of the formula~\cite{Moritz:Semant,MCAT,MLP,MLP:Improvement}. The Mathematical Language Processor (MLP) extracts definiens and identifier pairs based on context analyzation with NLP algorithms. 

During my first stay at NII, I researched on existing tools with the capability of parsing generic \LaTeX{} expressions, such as the POM-Tagger. We have chosen the content MathML (cMML) as the output format because the DLMF/DRMF macros are still not publicly available and thereby not supported by any other tools than the POM-Tagger and LaTeXML, which was originally developed to create the semantic version of the DLMF~\cite{LatexML}. We developed a comprehensive gold standard of 300 randomly picked formulae from Wikipedia, the DLMF sources and the NTCIR 12 task~\cite{MCAT,NTCIR12}. This gold standard contains manually annotated semantic information in manually corrected cMML files, such that the cMMLs can be considered as correct and most accurate for the 300 formulae in the gold standard. We compared the MML files generated from available tools with the cMML files in the gold standard to measure similarities and accuracy in extracting the identifiers and their meanings in the formula. Furthermore, we used the MLP to analyze the context of each formula from the gold standard. As a test state, we implemented some post-processing algorithms to improve the generated MML files with results from the MLP analyzation. For example, when the MLP recognizes a symbol as a function based on the context, we automatically adjust the generated MML based on this conclusion, which significantly improves the similarity and accuracy~\cite{JCDL:Paper}.

I will continue to research on the MLP pipeline and will combine the results from the context analyzation approach with the lexicon-based approach from the POM-Tagger. During our research, we miss a comprehensive API to handle MathML files effectively. Thereby, we are planning to provide such an API for others researchers that they do not need to reinvent the wheel and can follow our already beaten track. The next section will give a brief overview of the general objectives of my doctoral research.

\section{Thesis Objectives}
\subsection{Research Goal \& Research Objectives}
The goal of my doctoral research is to:
\begin{center}
\textit{Accomplish and evaluate an approach for automatic extraction of sufficient semantic information for calculations of mathematical expressions in scientific publications and web pages.}
\end{center}

In respect to this, we call semantic information of a mathematical expression sufficient when a translation to a CAS becomes feasible. To achieve this goal, I define the following thesis objectives:

\begin{enumerate}%
\item\label{o:I} Analyze the strengths and weaknesses of existing semantification approaches of mathematical expressions.
\item\label{o:II} Develop a new semantification concept that will improve the current approaches based on the identified weaknesses.
\item\label{o:III} Implement the system for an automatic semantification of mathematical expressions in real-world scientific documents.
\item\label{o:IV} Implement an extension of the system to provide translations to computer algebra systems.
\item\label{o:V} Evaluate the developed system by comparing the performance and accuracy, i.e., recall and precision values, with existing approaches and take advantage of evaluation techniques for CAS translations~\cite{MA}.
\end{enumerate}

In the following, I will give a brief overview of the already discovered problems of existing approaches and explain my new semantification concept.

\subsection{The Problems}
As already mentioned, an automatic semantification process has performance weaknesses and complicated to realize, because the semantics are rarely explicitly given in the expression. Consequently, one has to analyze the context of the expression to find and extract semantic information. While those extraction approaches usually already have high recall values, they still can be improved by improving the used NLP techniques, seen in the $\epsilon$-example from section~\ref{sec:relatedWork}. A possible improvement might take advantage of dictionary-based approaches. For example, consider a "\textit{constant}" as a noun in mathematical scientific publications. This would lead us to specialized NLP algorithms for scientific publications. Another problem is, that existing approaches also ignore the structure of mathematical expressions, which also contains useful information to conclude semantic information. Integrating these techniques as a preprocessing step and use specialized NLP algorithms would improve performances, recall and precision values. The following subsection will present my new approach that improves existing tools in the explained way.

\subsection{Multiple-Scan Approach}
Consider a dictionary, that contains information about the meanings of each symbol and the structure of the formula in respect of this meaning. The previously described POM-tagger project contains such a dictionary. Assuming the correct semantic information is given in that dictionary, we need to eliminate other possible meanings to conclude the correct information. My approach will take advantage of the existing tools and combine them with the mentioned dictionary-based approach. The concept will be split into the following three research objectives:

\begin{enumerate}
\item\label{o:1} narrow down possible meanings only from the expression itself, without referring to the context of the expression (dictionary-based),
\item\label{o:2} refine the process with conclusions from the nearby context of the expressions (dictionary-based preprocessing with context-based refinements), and 
\item\label{o:3} improve the previous process by analyzing not only the nearby context but the overall topic of the whole scientific paper or book, its references and other publications by the authors (further expanded context-based refinements).
\end{enumerate}

\begin{figure}[t]
	\centering
	\includegraphics[trim=1.3 1.3 1.3 1.3,clip,width=\textwidth]{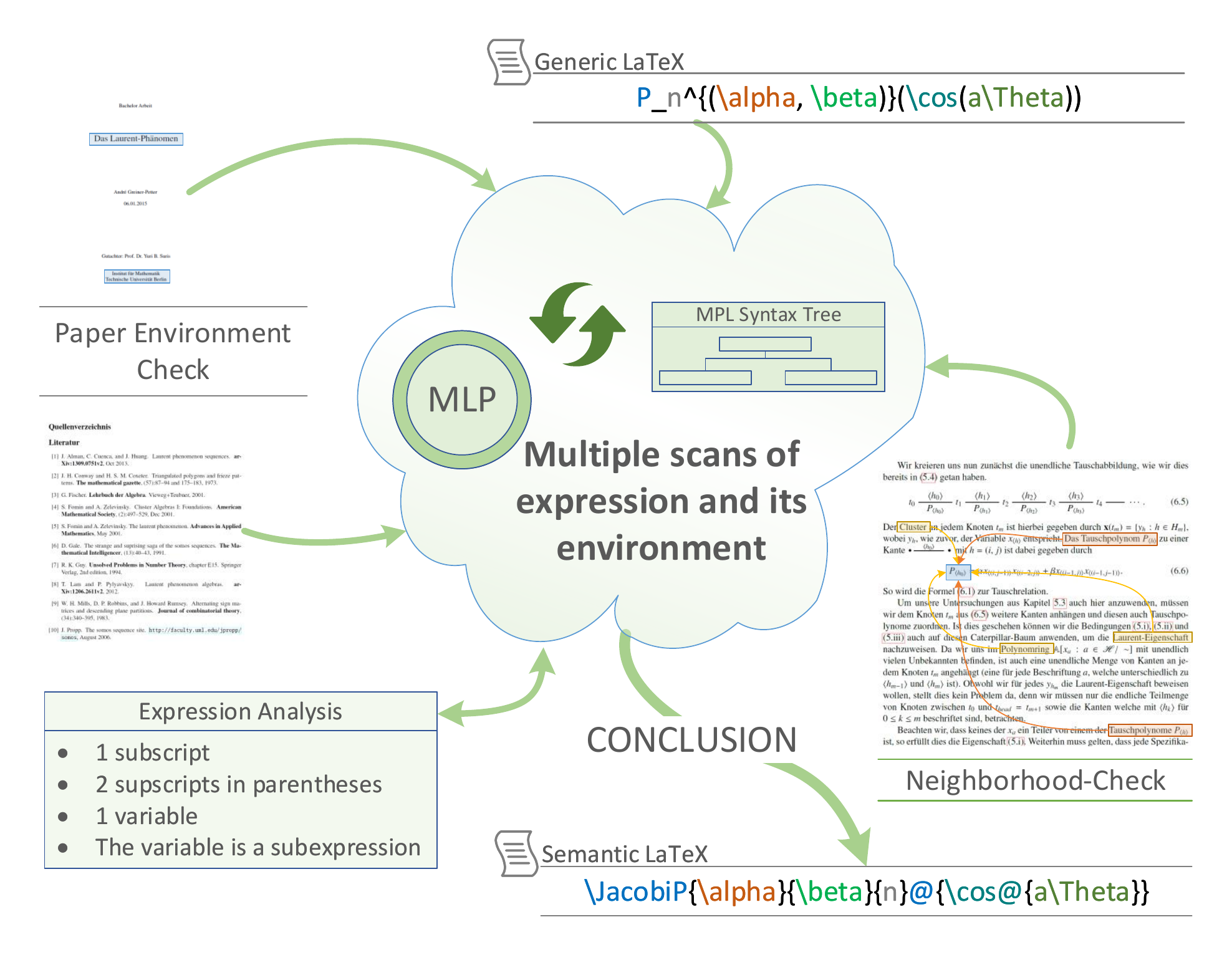}
	\caption{Conceptual explanation of the multiple scan approach.}
	\label{fig:multiScan}
\end{figure}

The behavior of a human reader inspires the idea behind the concept of the combination of context- and dictionary-based approaches shown in objectives~(\ref{o:1}), (\ref{o:2}) and (\ref{o:3}). The necessary information to get correct and essential semantics is given somewhere in the context and in the formula itself. If not, it would be difficult even for a scientific reader to understand the given expression. The concept is illustrated in figure~\ref{fig:multiScan}. In the following, I will explain each objective in a more detailed way.

Since we will achieve these three objectives by scanning the environment multiple times, we call this approach the \textit{multiple-scan approach}. Objective~(\ref{o:1}) concentrates on the expression itself, without extracting information from the context. My proposed approach is to exploit the coherence between the structure of given formula and its meaning, constructing a Markov logic network to deduce possible semantic meanings. Therefore, each meaning gets a probability. If the highest probability is below a given threshold, it would be necessary to use~(\ref{o:2}) and~(\ref{o:3}) for improving these probabilities. Otherwise, the probability is sufficiently high for concluding semantic information.

Consider the Jacobi polynomial expression in table~\ref{tab:useCase}. The given expression has a superscript, a subscript and the following expression in parentheses. A leading expression in letters with the following expression in parentheses may lead us to the conclusion that the leading expression is the name of a function and the expression in parentheses is its argument. That is also what Maple does, even when the name of the function is unknown. Additionally, the first symbol $P$ has a superscript and a subscript. Note that the Meixner-Pollaczek\footnote{Meixner-Pollaczek polynomial: $P_n^{(\lambda)}(x;\phi)$, \url{http://dlmf.nist.gov/18.19\#E6}} polynomial $P_n^{(\lambda)} (x;\phi)$ and the associated Legendre\footnote{Associated Legendre Function of the First Kind: $P_\nu^{\mu}(x)$, \url{http://dlmf.nist.gov/14.3\#E6}} function of the first kind $P_\nu^\mu (x)$ are also referenced with $P$ and all of these functions has a superscript, a subscript and an argument. But the Jacobi polynomial assumes a superscript of two parameters, while the Meixner-Pollaczek polynomial and the Legendre function just assume one parameter in the superscript.

Objective~(\ref{o:2}) based on the finding that around 70 percent of the symbolic elements in scientific papers are denoted in the surrounding text~\cite{MathWriting}. The mentioned NLP approaches try to exploit this finding to retrieve the semantic information for symbols in a formula. In my thesis, I will primarily work with the MLP approach described in~\cite{Moritz:Semant,MLP}. This approach extracts the symbols from the formula (called identifiers) and retrieves nouns from the surrounding text as candidates for definiens of the identifier. The scoring process assumes that the chance for a correct combination of identifier and definiens depends on the distance between the identifier and its definiens and the distance of the identifier to the closest formula that contains this identifier. I strongly believe we can improve the scoring process with the conclusions from my first objective above.

If the correct semantic information is still unsure, objective~(\ref{o:3}) is the last way to find a solution. Online compendia, such as arXiv, can be used to discover the overall topic of a scientific paper, the references and the area of research of the authors. The MCAT search engine developed by Kristianto, Topic, and Aizawa~\cite{MCAT,SemanticSearch} has performed such and can extract and score information from the document at the document granularity level. I will try to add this engine to our software to solve objective~(\ref{o:3}).

To achieve the objective~(\ref{o:1}), we will extend the POM-Tagger and its lexicon files. That is already part of the planned (and currently work in progress) development of the POM-Tagger. Therefore, I will focus on objectives~(\ref{o:2}) and~(\ref{o:3}), and support the progress of the POM-Tagger collaboratively with the DRMF project team. 

\section{Planned Research}
\subsection{Short Term Project Proposal}
Wikipedia as a highly frequently used lexicon has over 17 million edits every month\footnote{\url{https://stats.wikimedia.org/v2/\#/all-projects} (visited Feb. 2018)}. During the last two years (since May 2016 until January 2018) 7 million different formulae have been rendered via Mathoid in Wikipedia, i.e., there were 7 million edits during the last two years just in mathematical expressions\footnote{\url{https://github.com/physikerwelt/wikiMath17} (visited Feb. 2018)}. Wikipedia uses \TeX-Markup since 2003 to write and edit mathematical expressions. Therefore, the Wikipedia word processor is a highly suitable test environment to add machine learning algorithms.

The planned recommender system will be embedded to the Wikipedia word processor to learn and train supervised by each editor who modifies or write mathematical expressions in a Wikipedia article. The planned service would work as the usual word processor, that gets additional live updated recommended semantic versions for his mathematical \LaTeX{} input. An editor will be able to accept, refuse or just ignore the recommended semantic version for his current input. Based on his acceptance or refusal, the algorithm will adjust the scores for the recommended presentations.

Since every semantic macro has a generic \LaTeX{} definition, the backward translation dataset can be used for first training of the algorithm. Furthermore, the entire DLMF website is implemented in semantic \LaTeX{} and thus provide an eminently suitable test and training dataset.

\subsection{Long Term Plans}
As mentioned in the previous subsection, I plan to realize a machine learning algorithm integrated in Wikipedia. I estimate six months (Aug. 2018 - Jan. 2019) for developing the proposed project and additional three months (Feb. 2019 - Apr. 2019) for reviewing and evaluate the results. The results from this project will then be combined with the results from the previously developed and optimized MLP pipeline and the results from the first objective, which will be achieved by the POM-Tagger project.

The following task will be combining the results from the improved MLP project and the enhanced POM-Tagger project and develop an overall combination algorithm that takes advantage of both systems. This phase will finalize objectives~(\ref{o:1}) and~(\ref{o:2}) and realize a dictionary-based preprocessing system with context-based enhancements. I estimate three months for realizing this task (May 2019 - Jul. 2019).

With the help of the previously acquired findings, I will investigate the possibility of deducing semantic information from the overall topic of the context of an input expression. I am planning to use NLP approaches and online compendia of scientific papers such as arXiv to explore the topic by the authors and references of the scientific document. Accomplishing objective~(\ref{o:3}) will give us the field of a mathematical expression, so that we can fine-tune our scoring of objectives~(\ref{o:1}) and~(\ref{o:2}). I plan to dedicate six months (Aug. 2019 - Jan. 2020) to this task.

Writing the dissertation for a further six months (Jan. 2020 - Jun. 2020) will complete my doctoral research project.

\printbibliography[keyword=primary]
\end{document}